\documentstyle[twocolumn,aps,prl]{revtex}

\begin{document}


\title{Vortex Entanglement and Broken Symmetry}
\author{Andreas Sch\"onenberger$^{a}$, Vadim Geshkenbein$^{a,\, b}$, and Gianni
Blatter$^{a}$}
\address{$^{a\,}$Theoretische Physik, ETH-H\"onggerberg, CH-8093 Z\"urich,
Switzerland}
\address{$^{b\,}$L. D. Landau Institute for Theoretical Physics, 117940 Moscow,
Russia}

\twocolumn[
\date{\today}
\maketitle
\widetext
\vspace*{-1.0truecm}
\begin{abstract}
\begin{center}
\parbox{14cm}{
Based on the London approximation, we investigate numerically
the stability of the elementary configurations of entanglement,
the twisted-pair and the twisted-triplet, in the vortex-lattice
and -liquid phases. We find that, except for the dilute
limit, the twisted-pair is unstable and hence irrelevant in the
discussion of entanglement. In the lattice phase the twisted-triplet
constitutes a metastable, confined configuration of high energy.
Loss of lattice symmetry upon melting leads to deconfinement
and the twisted-triplet turns into a low-energy helical configuration.
}
\end{center}
\end{abstract}
]
\narrowtext

The combination of the soft elastic moduli and the large temperatures
attainable in the vortex system of the high-$T_c$ superconductors
boosts the importance of fluctuations and leads to interesting phenomena
such as vortex lattice melting and the appearance of vortex-liquid
phases\cite{review}. In
this context, topological excitations in the vortex system leading to
entanglement of the flux lines play an important role, both with respect to
statistical mechanics as well as dynamical properties of the vortex-solid and
-liquid phases. In this letter, we present a detailed analysis of the stability
and recombination properties of the elementary entangled configurations, the
twisted-pair and the twisted-triplet (see Fig. 1), for both the vortex-solid
and for a model vortex-liquid phase.

Topological excitations of the vortex-lattice in the form of edge- and
screw-dislocations are long time known objects\cite{Labusch}. Recently,
interest has
concentrated on more exotic configurations such as interstitials and
vacancies\cite{Frey}. The latter are relevant in the discussion
of a novel super-solid phase in layered high-$T_c$
superconductors\cite{Feigel,Frey}
and can be viewed as bound pairs of oppositely ``charged" edge
dislocations. Similarly, vortex entanglement is considered to be at the basis
of yet another new vortex phase, namely the vortex-liquid\cite{Nelson}, where
the
entanglement loops can be viewed as bound pairs of concentric screw
dislocations of opposite sign. The role played by such basic loops of
entanglement on the statistical mechanics of vortex-liquids is the
following\cite{review}:
If the equilibrium state of the vortex-liquid contains loops of all scales
with a finite density, the liquid exhibits a normal (dissipative) response
under application of a longitudinal current density ${\bf j}\parallel {\bf B}$,
and the entangled vortex-liquid is equivalent to the normal metallic
phase\cite{com1}.
If, on the other hand, the vortex-liquid remains disentangled, longitudinal
superconducting response survives the melting transition and we obtain a
new intermediate liquid phase distinct from the normal metallic one\cite{com2}.

Vortex entanglement is equally relevant for the dynamical properties of the
vortex-liquid: the prohibitively long relaxation times via
reptation\cite{oburubi} make the
barriers for vortex reswitching the limiting factor in the vortex
dynamics. The reswitching barriers then determine the inner viscosity of the
liquid and hence its pinning, creep, and flow properties\cite{review}.

Previous work on vortex cutting\cite{Wagen,Sudb} has
produced first estimates for the reswitching barrier of two isolated vortices;
however, as we will show below the relevance of these results for the
entanglement problem in vortex-solids and -liquids is rather
limited. Wilkin and Moore\cite{Wilm} have determined the excitation energy of
the crossing configuration of a vortex pair in a vortex-lattice, assuming this
configuration to constitute a saddle for vortex reswitching. However, as we
will show below, there is no metastable state for the twisted-pair
configuration in a vortex-solid and hence there exists no saddle for
reswitching (a similar result has recently been obtained by Dodgson and
Moore\cite{DM}
within the framework of the lowest Landau level approximation valid close to
$H_{c_2}$). A metastable twisted-pair configuration does exist in a model
vortex-liquid, where the pressure of the surrounding vortices acting on the
pair is modelled by a {\it circular} potential. This situation has been
investigated
by Carraro and Fisher\cite{Carraro}, who calculated the reswitching barrier for
the limiting
case of an infinitely extended twist using quite ingenious symmetry
arguments. Below we will argue that in a realistic description of the
vortex-fluid the
pressure excerted on a vortex pair by its surroundings destroys the metastable
twist in the
same way as in a vortex lattice and hence entanglements involving three or more
vortices have
to be considered.

Here, we study the stability and the reswitching barrier (connecting the
metastable entangled state with the stable rectilinear groundstate through a
saddle) for the elementary entangled configurations, the twisted-pair (TP) and
the
twisted-triplet (TT). The analysis is done for an isotropic superconductor
(penetration depth $\lambda$, coherence length $\xi$) within the London
approximation;
generalization of the results to the anisotropic situation involves simple
rescaling\cite{review}, at least for $B \gg H_{c_1}$. We start out with the
vortex-lattice and show that no metastable twisted-pair configuration exists
for
fields $B > H_{c_1}$. The TP state can be rendered metastable
either by artificially enhancing the vortex core energy or by going over to the
dilute limit $B \lesssim H_{c_1}$, where the interaction between vortices
becomes short
range. Next, we consider the twisted-triplet in the vortex-solid and find a
metastable state. The twist is restricted to a finite length
along the field axis and we call this a ``confined" excitation. The confinement
is a consequence of the discrete lattice symmetry and leads to a high
excitation energy when compared to the rectilinear groundstate. In comparison,
the
barrier stabilizing the TT state against reswitching is small and one
concludes that the lattice phase shows only little entanglement.
In the model vortex-liquid the situation remains essentially unchanged as
regards the twisted-pair --- the pressure of the neighboring vortices destroys
the metastable state in the same manner as in the vortex-solid phase.
For the twisted-triplet the assumption of a {\it
circular} effective potential mimicking the liquid environment is realistic.
The restoration of planar rotational symmetry upon melting leads to
deconfinement and
the TT turns into a low-energy helical configuration stabilized
by a high barrier against reswitching. As a consequence, one expects
the liquid to become entangled, however, due to the large reswitching
barrier, a non-entangled system only slowly transforms into an entangled state.
Whether a thermodynamic vortex-fluid phase becomes entangled immediately upon
melting
is a complicated statistical mechanics problem\cite{com2} and we will not go
into this
discussion here. In the following, we give a brief description of the
(numerical) technique on which our analysis is based and
then present the results for the twisted-pair and twisted-triplet
configurations in the vortex-lattice and the model vortex-liquid.

We base our analysis on the London approximation valid for the important low
and
intermediate field range $B < 0.2 \, H_{c_2}$. Choosing a set of $n$ vortices
(labelled
by $\mu, \nu = 1, \dots n$) involved in the excitation, the free energy
functional takes the form (see Ref.\cite{review}, ${\bf r}_\mu = ({\bf R}_\mu,
z))$
\begin{eqnarray*}
{\cal F}[{\bf r}_\mu] = \sum_{\mu = 1}^n \bigl\{{\cal F}_{{\rm self}}[{\bf
r}_\mu] +
{\cal F}_{{\rm surr}}[{\bf r}_\mu]\bigr\} + {1 \over 2} \sum_{\mu \neq \nu =
1}^n
{\cal F}_{{\rm int}}[{\bf r}_\mu, {\bf r}_\nu]
\end{eqnarray*}
with
\begin{eqnarray}
\nonumber
{\cal F}_{{\rm self}}&[&{\bf r}_\mu] = {\varepsilon_\circ \over 2} \int d{\bf
r}_\mu \cdot d{\bf
r}_{\mu'} \, { {e^{-\sqrt{{\bf r}_{\mu\mu'}^2 + \xi^2}/\lambda}} \over
{\sqrt{{\bf
r}_{\mu\mu'}^2 + \xi^2}} } + \varepsilon_\circ \int |d{\bf r}_\mu| \, c, \\
\nonumber
{\cal F}_{{\rm int}}&[&{\bf r}_\mu, {\bf r}_\nu] = \varepsilon_\circ \int d{\bf
r}_\mu
\cdot d{\bf r}_{\nu} \, { {e^{-\sqrt{{\bf r}_{\mu\nu}^2 + \xi^2}/\lambda}}
\over {\sqrt{{\bf
r}_{\mu\nu}^2 + \xi^2}} }, \\
\nonumber
{\cal F}_{{\rm surr}}&[&{\bf r}_\mu] = 2 \varepsilon_\circ \int d{\bf r}_\mu
\cdot {\bf
\hat z} \, V({\bf R}_\mu),
\end{eqnarray}
where ${\bf r}_\mu$ and ${\bf r}_{\mu'}$ (${\bf r}_{\mu\mu'} = {\bf r}_\mu -
{\bf
r}_{\mu'}$) refer to separate points on the same line. Here, we treat the
constant $c$ describing the vortex core energy as a parameter; within the
London
model its physical value is $c_{{\scriptscriptstyle 0}} \approx 0.5$.
Taking all nearest neighbors into account (see Fig. 1), we
choose $n = 10$ and $ n = 12$ for the TP and the TT, respectively. Twisted
metastable
configurations are obtained from a topologically correct initial state with
subsequent application of a conventional conjugate gradient method to minimize
the energy. In case a metastable state exists, the reswitching barrier is
found by imposing a constraint dragging the configuration from the
metastable minimum over the saddle towards the rectilinear groundstate.
Within a constrained configuration the vortex pair/triplet is forced to
have a prescribed distance $d_{{\scriptscriptstyle 0}}$ in the $z=0$ symmetry
plane.
With this constraint imposed, the metastable minimum is force-free, however,
the saddle
configuration in general remains forced and hence we obtain only an upper
estimate
for the energy of the true (force-free) saddle configuration.
We point out that neglecting the contribution of the core to
the self-energy ${\cal F}_{{\rm self}}$ produces severe instabilities leading
to unphysical
fluctuations upon relaxation. The origin of these short-range single-vortex
fluctuations can be traced back to the dispersive nature of the line tension
$\varepsilon_l (k_z) = \varepsilon_\circ {\rm ln} (1/k_z\xi)$. The functional
containing a finite core energy ($c_{{\scriptscriptstyle 0}} \approx 0.5$) is
not only physically
correct but also stable with respect to such pathological fluctuations.

Before turning to the specific discussion we briefly mention the natural scales
in the
problem: For a confined excitation the natural length scale along the field
axis is the
lattice constant $a_\circ$, whereas the scale for the excitation energy is
$\varepsilon_\circ a_\circ$. For the (uniaxially) anisotropic situation these
scales
change to $\varepsilon a_\circ$ for the length and
$\varepsilon\varepsilon_\circ a_\circ$
for the energy, where we have introduced the effective mass ratio
$\varepsilon^2 = m/M \ll
1$\cite{review}. The energy scale is conveniently expressed through the
melting temperature \cite{review}, $\varepsilon\varepsilon_\circ
a_\circ \approx T_{{\rm m}}/2.7
c_{{\scriptscriptstyle L}}^2 \approx 6 \, T_{{\rm m}}$, where we have used a
Lindemann
number $c_{{\scriptscriptstyle L}} = 0.25$ in the last equation.

{\it Vortex-lattice}: For the interaction with the surrounding vortices we
choose the
lattice potential
\begin{eqnarray*}
V({\bf R}) = \sum_{\bf m} K_{\scriptscriptstyle 0} (|{\bf R} - {\bf
R_m}|/\lambda) -
\sum_{\mu} K_{\scriptscriptstyle 0} (|{\bf R} - {\bf R}_\mu|/\lambda),
\end{eqnarray*}
with ${\bf R_m}$ denoting the equilibrium lattice sites and
$K_{\scriptscriptstyle 0}$
is the zero-order modified Bessel function. For intermediate fields
$H_{c_1} < B < H_{c_2}$ we can use the limit $\lambda \rightarrow \infty$ in
$V$,
leading to a simple and rapidly convergent series upon resummation in Fourier
space.
We then proceed along the lines described above and first search
for a metastable twisted-pair configuration. For fields $B > H_{c_1}$ and using
the
correct core energy with $c_{\scriptscriptstyle 0} \approx 0.5$ no such
metastable
state exists; in fact, although such a state does exist for the {\it isolated}
pair, it
is squeezed away by the pressure of the surrounding vortices as the lattice
potential $V$ is switched
on. Thereby the twisted vortices tend to align antiparallel in the
crossing region and the resulting {\it attractive} force between the segments
leads to their collapse and mutual annihilation. This tendency of antiparallel
alignment
can be suppressed by artificially increasing the vortex core energy:
for large enough $c$ such that $c > c_c \approx 4.9 - {\rm ln}(a_\circ/\xi)$
the metastable TP is recovered. Extrapolating this result (obtained for
$a_\circ/\xi < 20$)
to smaller fields, a metastable TP configuration is predicted for $a_\circ /
\xi >
\exp(4.9-c_{{\scriptscriptstyle 0}})\approx 80$ even in the limit $\lambda
\rightarrow
\infty$, a result relevant for $^4$He and marginally relevant for the
high-$T_c$
superconductors with $\kappa \approx 50$ -- 100.
A (more physical) alternative to stabilize the
twisted-pair configuration is to decrease the surrounding pressure via reducing
the vortex density. For fields $B \lesssim H_{c_1}$ the interaction becomes
short-range and we have to return to a finite screening length $\lambda <
\infty$
in $V$. Indeed, in the dilute limit we find a metastable TP for fields
$B < 1.65\, H_{c_1}$, {\it i.\ e.,} $a_\circ > 1.8 \,\lambda$, where $\kappa =
10$ has been
chosen.

Next we discuss the twisted-triplet state in the vortex-lattice (results for
$a_\circ / \xi = 10$ are quoted in the text; see Table I for a summary).
Following the scheme described above we find a metastable TT
configuration well in the London regime, {\it i.\ e.,} with a minimal
separation
$d_{{\rm min}} \approx 0.8 \, a_\circ \gg \xi$ between the vortices. Due to the
hexagonal
symmetry of the surrounding lattice, the twist is constrained within a length
$L_{{\scriptscriptstyle TT}} \approx 2.2 \, a_\circ$ and its energy is high as
compared to the
rectilinear groundstate, $E_{{\scriptscriptstyle TT}} - E_{{\scriptscriptstyle
0}}
\approx 2.1 \, \varepsilon_\circ a_\circ$, see Fig. 2.
On the other hand, the barrier stabilizing the
metastable state is small, $E_s - E_{{\scriptscriptstyle TT}} \approx 0.32 \,
\varepsilon_\circ a_\circ$. The saddle still is in the London regime with
$d_{{\rm min}} > 0.3 \, a_\circ$. The reswitching goes through a first
collapse, leading
to the creation of a small transverse loop which subsquently shrinks to zero,
leaving
behind three rectilinear vortices in their groundstate, see the inset of Fig.
2.
Note the sharp drop in energy when crossing the saddle, a consequence of the
sudden reswitching and an indication that the present constraint does not
produce a
force-free configuration at the saddle-point.
We defer the detailed discussion of the various types of (re)switching saddles
(loop
creation, hysteretic effects) to a forthcoming paper.

{\it Model vortex-liquid:} Here we reduce the description to those
vortices directly involved in the entanglement, {\it i.\ e.,} $n = 2$ and
$n = 3$ for the twisted-pair and twisted-triplet, respectively. We choose the
surrounding potential mimicking the vortex-liquid to be of the form
\begin{eqnarray*}
V({\bf R}) = {f \over 2} (x^2 + s^2 y^2),
\end{eqnarray*}
with $f = 2/a_\circ^2$ ($f = 3/a_\circ^2$) for the TP (TT) configuration;
$s$ denotes an asymmetry parameter. In their analysis of the TP problem,
Carraro and
Fisher\cite{Carraro} chose a circular symmetry with $s=1$, where the metastable
twisted state takes a helical shape (with a pitch $L_p$) and is degenerate with
the
rectilinear groundstate in the limit $L_p \rightarrow \infty$. In this limit,
the
reswitching barrier is characterized by a high degree of symmetry and using
$a_\circ / \xi =
10$ we find $E_s - E_h \approx 1.0\, \varepsilon_\circ a_\circ$, in agreement
with previous
work\cite{Carraro}. However, in a
liquid phase one does not expect the potential acting on a vortex pair to have
a circular
symmetry; rather the pressure of the surrounding vortices produces an
asymmetric potential
with $s > 1$ (one expects $s \approx 2.7$ on simple geometrical grounds,
assuming a similar pressure to act in the liquid phase as in the solid one).
Increasing
$s$ beyond the critical value $s_c \approx 1.3$ the entangled configuration
becomes unstable and
no metastable TP configuration is expected to exist in the liquid phase either.
One could argue, that in a vortex liquid the asymmetric potential can
accommodate itself and
rotate along with the pair, however, such a configuration involves the twist of
four lines at
least. The approximation of a circular effective potential mimicking the
pressure of the
surrounding liquid is already quite reasonable for the triplet (and further
improves for the
configurations involving larger loops with 6 and more vortices). In fact, the
main reason for
studying the triplet as the elementary entangled configuration is the existence
of a
metastable twisted configuration in the solid, implying the existence of a
similar metastable
state in the liquid which still exhibits lattice order on short spatial and
temporal scales; in a
(viscous) liquid, the entanglement will first go through a confined high-energy
twist, similar
to the one in the lattice, which subsequently relaxes into a low-energy helical
state due to
the absence of shear forces in the liquid. The excitation energy of this
relaxed state
depends on the pitch $L_p$ of the helix and we have verified that for $L_p \gg
a_\circ$
the tilt energy $E_h - E_{{\scriptscriptstyle 0}} \approx (2 \pi^2/9)
[{\rm ln} (a_\circ/\xi) + 0.5] \varepsilon_\circ a_\circ^2/L_p$
provides a good approximation within the London regime (note that here the
vortices twist by
the angle $2\pi/3$ on the length $L_p$). The saddle-point energy for
(re)switching remains
high, however; in the limit $L_p \rightarrow \infty$ an analysis based on
symmetry arguments
similar to those used by Carraro and Fisher\cite{Carraro} provides the result
$E_s - E_h
\approx 2.4 \, \varepsilon_\circ a_\circ$, where $a_\circ/\xi = 10$ and
$\lambda \rightarrow
\infty$ has been chosen. Results for other values of $a_\circ / \xi$ are
summarized in Table I.

In conclusion, we have investigated the elementary entangled configurations,
the twisted-pair and the twisted-triplet in a vortex-lattice and -liquid phase.
Away from the dilute limit, we find that the twisted-pair is unstable and
hence irrelevant for the discussion of entanglement. The basic loop of
entanglement
is the twisted-triplet which is metastable both in the vortex-lattice and
-liquid phase.
In the vortex-lattice, the TT is a high energy state stabilized by a
comparatively
small barrier and hence the vortex-lattice does not entangle. The high energy
of
the TT excitation is a consequence of the confinement produced by the lattice
symmetry of the surrounding potential. Upon melting, the shear modulus
disappears and
the hexagonal lattice symmetry changes to a rotational one in the liquid.
The high-energy twisted-triplet dissolves into a low-energy helix with a
pitch $L_p$ determined by the mutual entanglement in the liquid.
It is the phenomenon of deconfinement and its associated drop in excitation
energy which tends to bind the two (originally unrelated) transitions of
melting (loss of translational lattice symmetry) and entanglement (loss of
longitudinal superconductivity) together. Whereas melting {\it immediately}
triggers
entanglement in a system with short range interactions, it remains to be shown
whether a disentangled liquid state can be stabilized in a system characterized
by an
interaction with long range.

We thank M. Feigel'man, E. Heeb, A. Larkin, H. Nordborg, and A. van Otterlo,
for interesting
and useful discussions. Financial support from the Fonds National Suisse and
the
International Soros Foundation (grant \# M6M000) is gratefully acknowledged.

\begin{table}
\caption{
Numerical results obtained for the triplet configurations in the vortex-solid
(s)
and -liquid (l) phases.
}

\begin{tabular}{dddd}
$a_\circ/\xi$ & $L_{{\scriptscriptstyle TT},p}/a_\circ$ &
      $(E_{{\scriptscriptstyle TT},h}-E_{{\scriptscriptstyle
0}})/\varepsilon_\circ a_\circ$ &
      $(E_s-E_{{\scriptscriptstyle TT},h})/\varepsilon_\circ a_\circ$ \\
\hline
 5 (s)& 1.7 & 1.62 & 0.05\\
 5 (l)& $\infty$ & 0 & 1.6\\
10 (s)& 2.2 & 2.12& 0.32\\
10 (l)& $\infty$ & 0 & 2.4\\
20 (s)& 2.7 & 2.42 & $\sim$ 1.2\\
20 (l)& $\infty$ & 0 & 3.2\\
\end{tabular}
\end{table}

\begin{figure}
\caption{Twisted-triplet configuration embedded within a vortex-lattice. The 9
nearest
neighbors are allowed to relax when twisting the inner vortex triplet. A
confined
($L_{{\scriptscriptstyle TT}} \approx 2.2 \, a_\circ$), high
energy metastable state is found, with an excitation energy
$E_{{\scriptscriptstyle TT}}
- E_{{\scriptscriptstyle 0}} \approx 2.12\, \varepsilon_\circ a_\circ$ and
stabilized
against reswitching by the small barrier $E_s - E_{{\scriptscriptstyle TT}}
\approx 0.32\,
\varepsilon_\circ a_\circ$.
}
\label{fig:1}
\end{figure}

\begin{figure}
\caption{
Excitation energy $E_{{\scriptscriptstyle TT}} (d_{{\scriptscriptstyle 0}}) -
E_{{\scriptscriptstyle 0}}$ versus distance $d_{{\scriptscriptstyle 0}}$ for
the
clamped twisted-triplet configuration in a vortex-lattice. Inset shows top-view
of
the clamped configurations marked by the arrows, particularly the metastable
state and the
fast reswitching geometry close to the saddle. Note the small relaxation
amplitude of the
surrounding vortices. Upon melting, the confined high-energy twisted
configuration turns
into a low-energy helical state, whereas the saddle-point configuration for
(re)switching
stays confined and remains at high energy (see Table I).
}
\label{fig:2}
\end{figure}


\begin{thebibliography}{99}
\bibitem{review} G. Blatter, M.\ V.\ Feigel'man, V.\ B.\ Geshkenbein, A.\ I.\
Larkin,
and V.\ M.\ Vinokur,
{\sl Rev. Mod. Phys.} {\bf 6x}, yyyy (1995).
\bibitem{Labusch} R.\ Labusch, {\sl Phys. Lett.} {\bf 22}, 9 (1966).
\bibitem{Frey} E.\ Frey, D.\ R.\ Nelson, and D.\ S.\ Fisher, {\sl Phys. Rev. B}
{\bf 49},
9723 (1994).
\bibitem{Feigel} M.\ V.\ Feigel'man, V.\ B.\ Geshkenbein, and A.\ I.\ Larkin,
{\sl
Physica C} {\bf 167}, 177 (1990).
\bibitem{Nelson} D.\ R.\ Nelson, {\sl Phys. Rev. Lett.} {\bf 60}, 1973 (1988).
\bibitem{com1} This phase corresponds to the superfluid groundstate of bosons
when
mapping the 3D classical vortex problem to a 2D quantum particle problem; the
entanglement
loops correspond to cooperative ring exchanges producing finite winding
number and hence superfluidity in the Bose-system.
\bibitem{com2} The existence of this phase has been proposed by Feigel'man
({\sl Physica A} {\bf 168}, 319 (1990), see also Ref. [1])
and corresponds to a new {\it normal} groundstate in a system of charged 2D
bosons.
\bibitem{oburubi} S.\ P.\ Obukhov and M.\ Rubinstein, {\sl Phys. Rev. Lett.}
{\bf 65},
1279 (1990).
\bibitem{Wagen} P.\ Wagenleithner, {\sl J. Low. Temp. Phys.} {\bf 48}, 25
(1982).
\bibitem{Sudb} A.\ Sudb{\o} and E.\ H.\ Brandt, {\sl Phys. Rev. Lett.} {\bf
67}, 3176
(1991).
\bibitem{Wilm} N.\ K.\ Wilkin and M.\ A.\ Moore, {\sl Phys. Rev. B} {\bf 48},
3464
(1993).
\bibitem{DM} M.\ J.\ W.\ Dodgson and M.\ A.\ Moore, {\sl preprint}, University
of
Manchester (1994).
\bibitem{Carraro} C.\ Carraro and D. S. Fisher, {\sl preprint}, Harvard
University
(1994).
\end{thebibliography}
\end{document}